# Even Good Bots Fight: The Case of Wikipedia

Milena Tsvetkova [a], Ruth García-Gavilanes [a], Luciano Floridi [a,b], and Taha Yasseri [a,b*]

[a] Oxford Internet Institute, University of Oxford, Oxford OX1 3JS, United Kingdom

[b] Alan Turing Institute, London NW1 2DB, United Kingdom

* Corresponding author. E-mail: taha.yasseri@oii.ox.ac.uk (TY)

## Abstract

In recent years, there has been a huge increase in the number of bots online, varying from Web crawlers for search engines, to chatbots for online customer service, spambots on social media, and content-editing bots in online collaboration communities. The online world has turned into an ecosystem of bots. However, our knowledge of how these automated agents are interacting with each other is rather poor. Bots are predictable automatons that do not have the capacity for emotions, meaning-making, creativity, and sociality and it is hence natural to expect interactions between bots to be relatively predictable and uneventful. In this article, we analyze the interactions between bots that edit articles on Wikipedia. We track the extent to which bots undid each other's edits over the period 2001-2010, model how pairs of bots interact over time, and identify different types of interaction trajectories. We find that, although Wikipedia bots are intended to support the encyclopedia, they often undo each other's edits and these sterile "fights" may sometimes continue for years. Unlike humans on Wikipedia, bots' interactions tend to occur over longer periods of time and to be more reciprocated. Yet, just like humans, bots in different cultural environments may behave differently. Our research suggests that even relatively "dumb" bots may give rise to complex interactions, and this carries important implications for Artificial Intelligence research. Understanding what affects bot-bot interactions is crucial for managing social media well, providing adequate cyber-security, and designing well functioning autonomous vehicles.

# Introduction

In August 2011, Igor Labutov and Jason Yosinski, two PhD students at Cornell University, let a pair of chat bots, called Alan and Sruthi, talk to each other online. Starting with a simple greeting, the one-and-a-half-minute dialogue quickly escalated into an argument about what Alan and Sruthi had just said, whether they were robots, and about God (1). The first ever conversation between two simple artificial intelligence agents ended in a conflict.

A bot, or software agent, is a computer program that is persistent, autonomous, and reactive (2,3). Bots are defined by programming code that runs continuously and can be activated by itself. They make and execute decisions without human intervention and perceive and adapt to the context they operate in. Internet bots, also known as web bots, are bots that run over the Internet. They appeared and proliferated soon after the creation of the World Wide Web (4). Already in 1993, Martijn Koster published "Guidelines to robot writers," which contained suggestions about developing web crawlers (5), a kind of bot. Eggdrop, one of the first known Internet Relay Chat bots, started greeting chat newcomers also in 1993 (6). In 1996, Fah-Chun Cheong published a 413-page book, claiming to have a current listing of all bots available on the Internet at that point in time. Since then, Internet bots have proliferated and diversified well beyond our ability to record them in an exhaustive list (7,8). As a result, bots have been responsible for an increasingly larger proportion of activities on the Web. For example, one study found that 25% of all messages on Yahoo! chat over a period of three months in 2007 were sent by spam bots (9). Another study discovered that 32% of all tweets made by the most active Twitter users in 2009 were generated by bots (10), meaning that bots were responsible for an estimated 24% of all tweets (11). Further, researchers estimated that bots comprise between 4% and 7% of the avatars on the virtual world Second Life in 2009 (12). A media analytics company found that 54% of the online ads shown in thousands of ad campaigns in 2012 and 2013 were viewed by bots, rather than humans (13). According to an online security company, bots accounted for 48.5% of website visits in 2015 (14). Also in 2015, 100,000 accounts on the multi-player online game World of Warcraft (about 1% of all accounts) were banned for using bots (15). And in the same year, a database leak revealed that more than 70,000 "female" bots sent more than 20 million messages on the cheater dating site Ashley Madison (16).

**Table 1. Categorization of Internet bots according to the intended effect of their operations and the kind of activities they perform, including some familiar examples for each type.**

| | **Benevolent** | **Malevolent** |
|---|---|---|
| **Collect information** | • Web crawlers<br>• Bots used by researchers | • Spam bots that collect e-mail addresses<br>• Facebook bots that collect private information |
| **Execute actions** | • Anti-vandalism bots on Wikipedia<br>• Censoring and moderating bots on chats and forums | • Auction-site bots<br>• High-frequency trading algorithms<br>• Gaming bots<br>• DDoS attack bots<br>• Viruses and worms<br>• Clickfraud bots that increase views of online ads and YouTube videos |
| **Generate content** | • Editing bots on Wikipedia<br>• Twitter bots that create alerts or provide content aggregation | • Spam bots that disseminate ads<br>• Bot farms that write positive reviews and boost ratings on Apple App Store, YouTube, etc. |
| **Emulate humans** | • Customer service bots<br>• @DeepDrumpf and poet-writing bots on Twitter<br>• AI bots, e.g. IBM's Watson | • Social bots involved in astroturfing on Twitter<br>• Social bots on the cheater dating site Ashley Madison |

Benevolent bots are designed to support human users or cooperate with them. Malevolent bots are designed to exploit human users and compete negatively with them. We have classified high-frequency trading algorithms as malevolent because they exploit markets in ways that increase volatility and precipitate flash crashes. In this study, we use data from editing bots on Wikipedia (benevolent bots that generate content).

As the population of bots active on the Internet 24/7 is growing fast, their interactions are equally intensifying. An increasing number of decisions, options, choices, and services depend now on bots working properly, efficaciously, and successfully. Yet, we know very little about the life and evolution of our digital minions. In particular, predicting how bots' interactions will evolve and play out even when they rely on very simple algorithms is already challenging. Furthermore, as Alan and Sruthi demonstrated, even if bots are designed to collaborate, conflict may occur inadvertently. Clearly, it is crucial to understand what could affect bot-bot interactions in order to design cooperative bots that can manage disagreement, avoid unproductive conflict, and fulfill their tasks in ways that are socially and ethically acceptable.

There are many types of Internet bots (see Table 1). These bots form an increasingly complex system of social interactions. Do bots interact with each other in ways that are comparable to how we humans interact with each other? Bots are predictable automatons that do not have the capacity for emotions, meaning-making, creativity, and sociality (17). Despite recent advances in the field of Artificial Intelligence, the idea that bots can have morality and culture is still far from reality. Today, it is natural to expect interactions between bots to be relatively predictable and uneventful, lacking the spontaneity and complexity of human social interactions. However, even in such simple contexts, our research shows that there may be more similarities between bots and humans than one may expect. Focusing on one particular human-bot community, we find that conflict emerges even among benevolent bots that are designed to benefit their environment and not fight each other, and that bot interactions may differ when they occur in environments influenced by different human cultures.

We study bots on Wikipedia, the largest free online encyclopedia. Bots on Wikipedia are computer scripts that automatically handle repetitive and mundane tasks to develop, improve, and maintain the encyclopedia. They are easy to identify because they operate from dedicated user accounts that have been flagged and officially approved. Approval requires that the bot follows Wikipedia's bot policy.

Bots are important contributors to Wikipedia. For example, in 2014, bots completed about 15% of the edits on all language editions of the encyclopedia (18). In general, Wikipedia bots complete a variety of activities. They identify and undo vandalism, enforce bans, check spelling, create inter-language links, import content automatically, mine data, identify copyright violations, greet newcomers, and so on (19). Our analysis here focuses on editing bots, which modify articles directly. We analyze the interactions between bots and investigate the extent to which they resemble interactions between humans. In particular, we focus on whether bots disagree with each other, how the dynamics of disagreement differ for bots versus humans, and whether there are differences between bots operating in different language editions of Wikipedia.

To measure disagreement, we study reverts. A revert on Wikipedia occurs when an editor, whether human or bot, undoes another editor's contribution by restoring an earlier version of the article. Reverts that occur systematically indicate controversy and conflict (20–22). Reverts are technically easy to detect regardless of the context and the language, so they enable analysis at the scale of the whole system.

Our data contain all edits in 13 different language editions of Wikipedia in the first ten years after the encyclopedia was launched (2001-2010). The languages represent editions of different size and editors from diverse cultures (see Materials and Methods for details). We know which user completed the edit, when, in which article, whether the edit was a revert and, if so, which previous edit was reverted. We first identified which editors are humans, bots, or vandals. We isolated the vandals since their short-lived disruptive activity exhibits different time and interaction patterns than the activity of regular Wikipedia editors.

# Results

Bots constitute a tiny proportion of all Wikipedia editors but they stand behind a significant proportion of all edits (Figs 1A and 1B). There are significant differences between different languages in terms of how active bots are. From previous research, we know that, in small and endangered languages, bots are extremely active and do more than 50% of the edits, sometimes up to 100% (19). Their tasks, however, are mainly restricted to adding links between articles and languages. In large and active languages, the level of bot activity is much lower but also much more variable.

Compared to humans, a smaller proportion of bots' edits are reverts and a smaller proportion get reverted (Figs 1C and 1D). In other words, bots dispute others and are disputed by others to a lesser extent than humans. Since 2001, the number of bots and their activity has been increasing but at a slowing rate (S1 Fig). In contrast, the number of reverts between bots has been continuously increasing (Fig 2A). This would suggest that bot interactions are not becoming more efficient. We also see that the proportion of mutual bot-bot reverts has remained relatively stable, perhaps even slightly increasing over time, indicating that bot owners have not learned to identify bot conflicts faster (Fig 2B).

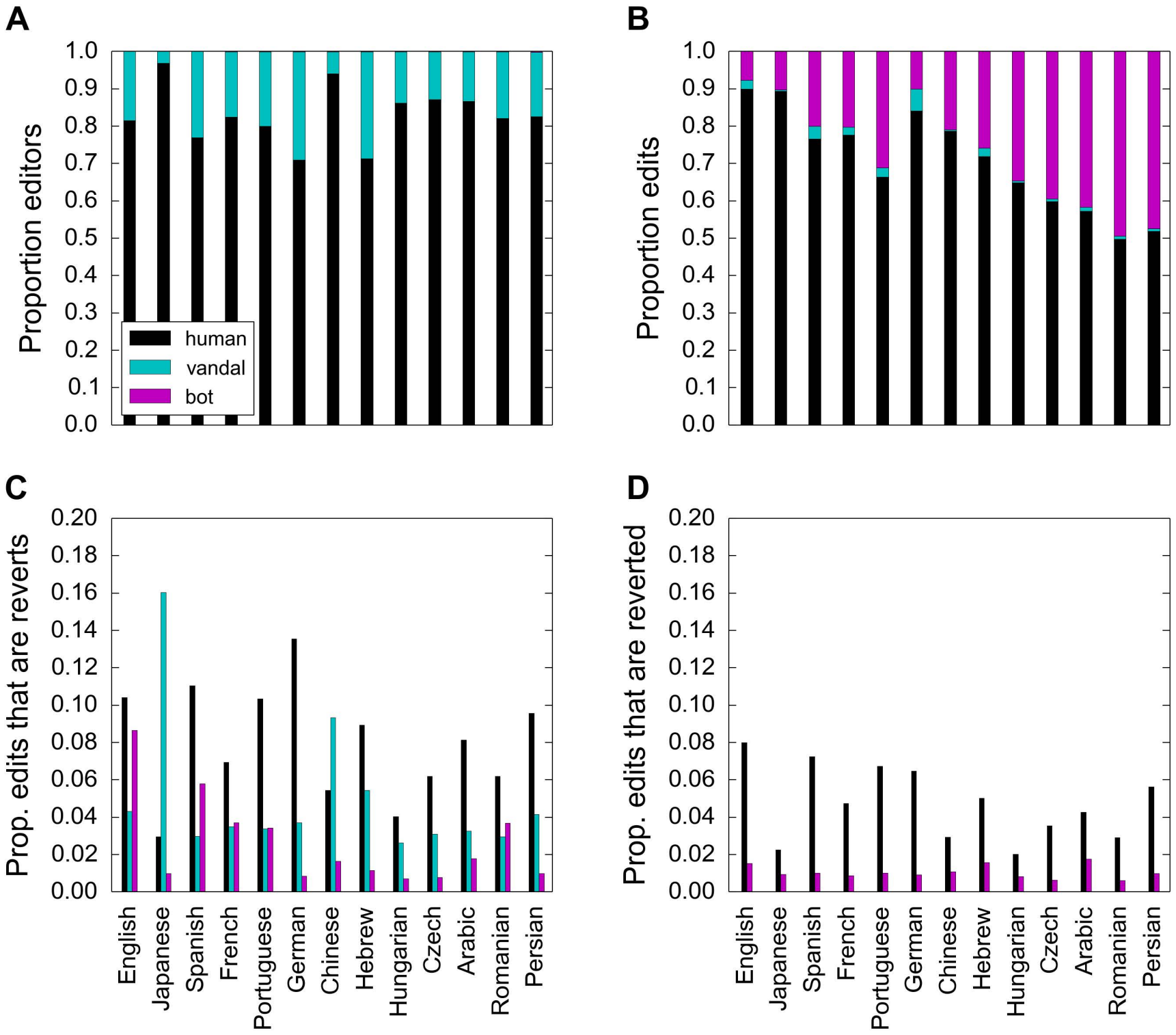

**Fig 1. The proportion of Wikipedia editors who are human, vandals, and bots and the type of editorial activity in which they are involved.** A language edition to the left has a higher total number of edits than one to the right. (A) Bots comprise a tiny proportion of all Wikipedia users, usually less than 0.1% (not visible in the figure). (B) However, bots account for a significant proportion of the editorial activity. The level of bot activity significantly differs between different language editions of Wikipedia, with bots generally more active in smaller editions. (C) A smaller proportion of bots' edits are reverts compared to humans' edits. (D) A smaller proportion of bots' edits get reverted compared to humans' edits. Since by our definition, vandals have all of their edits reverted, we do not show them in this figure.

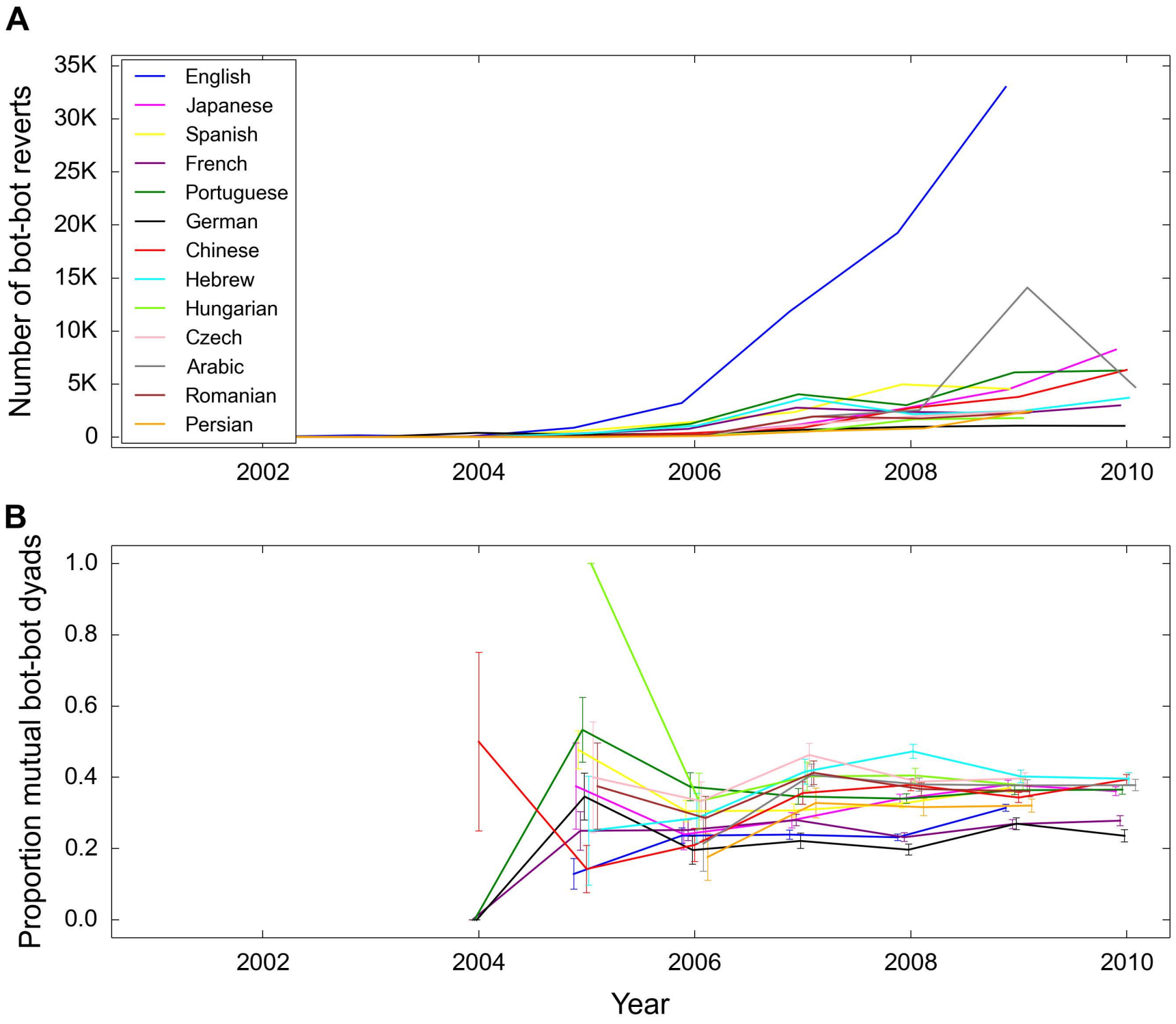

**Fig 2. The number of bot reverts executed by another bot and the proportion of unique bot-bot pairs that have at least one reciprocated revert for the period 2001-2010.** (A) Generally, the number of bot-bot reverts has been increasing. (B) However, the proportion of reciprocated reverts has not been decreasing (error bars correspond to one standard error). This suggests that disagreement between bots is not becoming less common.

In general, bots revert each other a lot: for example, over the ten-year period, bots on English Wikipedia reverted another bot on average 105 times, which is significantly larger than the average of 3 times for humans (S1 Table). Bots on German Wikipedia revert each other to a much lesser extent than other bots (24 times on average). Bots on Portuguese Wikipedia, in contrast, fight the most, with an average of 185 bot-bot reverts per bot. This striking difference, however, disappears when we account for the fact that bots on Portuguese Wikipedia edit more than bots on German Wikipedia. In general, since bots are much more active editors than humans, the higher number of bot-bot reverts does not mean that bots fight more than humans. In fact, the proportion of bots' edits that are reverts is smaller for bots than for humans (Fig 1C). This proportion is highest for bots in the English and the Romance-language editions (Spanish, French, Portuguese, and Romanian). Interestingly, although bots in these languages revert more often compared to bots in other languages, fewer of these reverts are for another bot (S2 Fig).

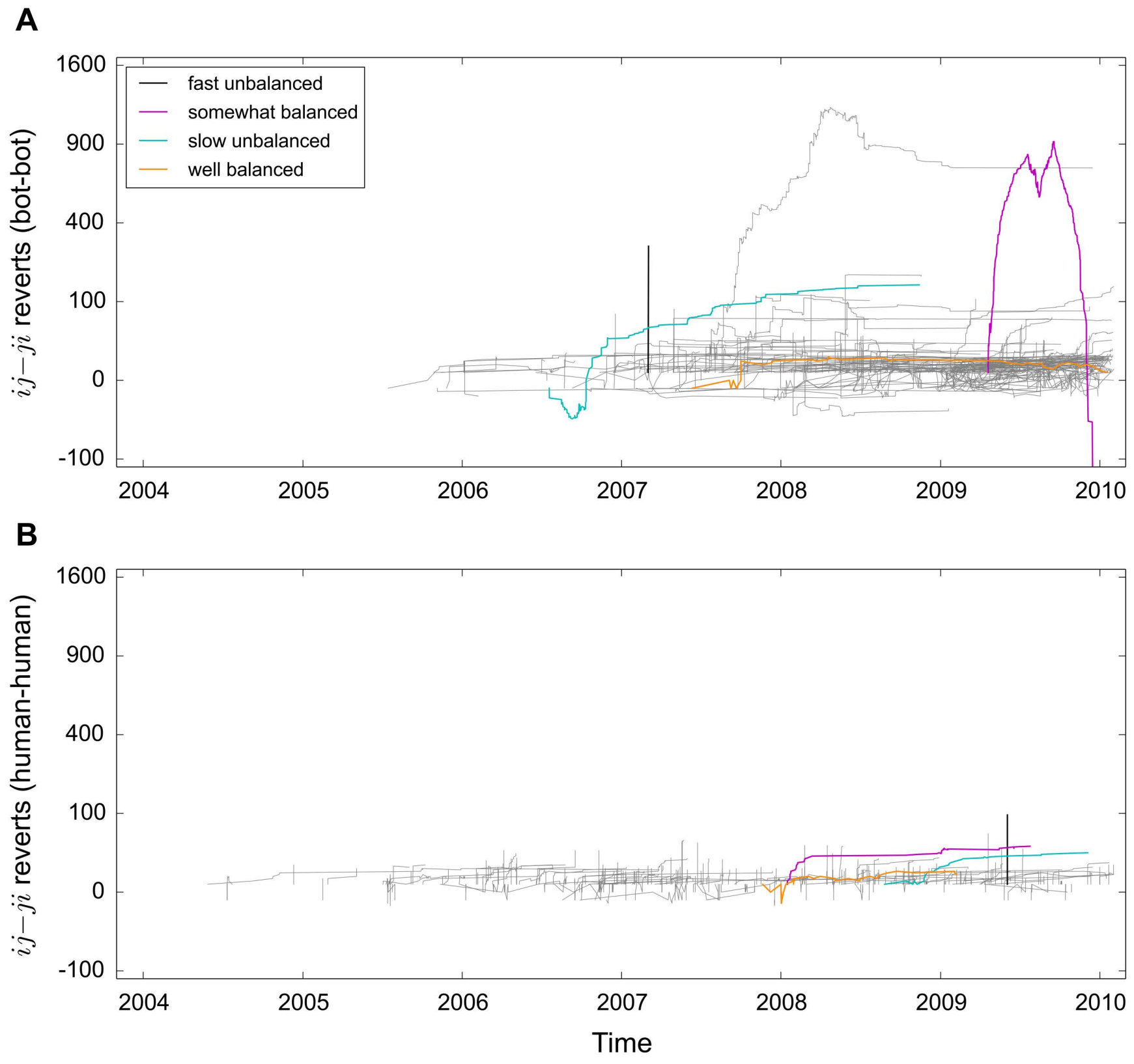

**Fig 3. Typical interaction trajectories for bot-bot and human-human pairs in English Wikipedia in the period 2001-2010.** The interaction trajectories are constructed as follows: starting from $y_o = 0$, $y_t = y_{t-1} + 1$ if $i$ reverts $j$ and $y_t = y_{t-1} - 1$ if $j$ reverts $i$ at time $t$; the labels $i$ and $j$ are assigned so that $y >= 0$ for the majority of the $ij$ interaction time; to compress the extremes, we scaled the y-axis to the power of 0.5. The panels show the trajectories of 200 pairs randomly sampled from those who have exchanged more than five reverts. In addition, we highlight the four longest trajectories in the sample from each of the four trajectory types we identify. Compared to human-human interactions, bot-bot interactions occur at a slower rate and are more balanced, in the sense that reverts go back and forth between the two editors.

Our analysis focuses on interactions in dyads over time. We model the interaction trajectories in two-dimensional space, where the x-axis measures time and the y-axis measures how many more times the first editor has reverted the second compared to the second reverting the first (Fig 3). We analyze three properties of the trajectories: latency, imbalance, and reciprocity. Latency measures the average steepness of the interaction trajectory, imbalance measures the distance between the *x*-axis and the last point of the trajectory, and reciprocity measures the trajectory's jaggedness (see Materials and Methods below for definitions).

Analyzing the properties of the interaction trajectories suggests that the dynamics of disagreement differ significantly between bots and humans. Reverts between bots tend to occur at a slower rate and a conflict between two bots can take place over longer periods of time, sometimes over years. In fact, bot-bot interactions have different characteristic time scale than human-human interactions (S3 Fig). The characteristic average time between successive reverts for humans is at 2 minutes, 24 hours, or 1 year. In comparison, bot-bot interactions have a characteristic average response of 1 month. This difference is likely because, first, bots systematically crawl articles and, second, bots are restricted as to how often they can make edits (the Wikipedia bot policy usually requires spacing of 10 seconds, or 5 for anti-vandalism activity, which is considered more urgent). In contrast, humans use automatic tools that report live changes made to a pre-selected list of articles (24,25); they can thus follow only a small set of articles and, in principle, react instantaneously to any edits on those.

Bots also tend to reciprocate each other's reverts to a greater extent. In contrast, humans tend to have highly unbalanced interactions, where one individual unilaterally reverts another one (S4 and S5 Figs).

We quantify these findings more precisely by identifying different types of interaction trajectories and counting how often they occur for bots and for humans, as well as for specific languages. To this end, we use k-means clustering on the three properties of the trajectories (latency, imbalance, and reciprocity) and on all bot-bot and human-human interactions longer than five reverts (the results are substantively similar without the length restriction). We do not claim that the clusters are natural to the data; rather, we use the clusters to compare the interactions of the different groups.

The algorithm suggested that the data can be best clustered in four trajectory types (S6 Fig):

- **Fast unbalanced trajectories**. These trajectories have low reciprocity and latency and high imbalance. They look like smooth vertical lines above the x-axis.
- **Slow unbalanced trajectories**. These trajectories have low reciprocity and high latency and imbalance. They look like smooth diagonal lines above the x-axis.
- **Somewhat balanced trajectories**. These trajectories have intermediate imbalance and reciprocity. They are somewhat jagged and cross the x-axis.
- **Well balanced trajectories**. These trajectories have low imbalance and high reciprocity. They are quite jagged and centered on the x-axis.

| | Bot-bot | | | | Human-human | | | |
|---|---|---|---|---|---|---|---|---|
| | Fast unbalanced | Slow unbalanced | Somewhat balanced | Well balanced | Fast unbalanced | Slow unbalanced | Somewhat balanced | Well balanced |
| English | 0.20 | 0.28 | 0.40 | 0.12 | 0.43 | 0.14 | 0.19 | 0.23 |
| Japanese | 0.03 | 0.31 | 0.44 | 0.23 | 0.51 | 0.15 | 0.23 | 0.12 |
| Spanish | 0.09 | 0.24 | 0.46 | 0.22 | 0.50 | 0.17 | 0.17 | 0.16 |
| Portuguese | 0.08 | 0.33 | 0.40 | 0.20 | 0.42 | 0.22 | 0.18 | 0.18 |
| French | 0.14 | 0.28 | 0.39 | 0.20 | 0.52 | 0.11 | 0.16 | 0.21 |
| Chinese | 0.04 | 0.29 | 0.45 | 0.22 | 0.40 | 0.16 | 0.26 | 0.18 |
| German | 0.11 | 0.26 | 0.42 | 0.22 | 0.33 | 0.10 | 0.19 | 0.37 |
| Hebrew | 0.04 | 0.23 | 0.43 | 0.29 | 0.34 | 0.23 | 0.22 | 0.21 |
| Hungarian | 0.09 | 0.22 | 0.38 | 0.31 | 0.43 | 0.17 | 0.18 | 0.21 |
| Arabic | 0.04 | 0.27 | 0.42 | 0.28 | 0.34 | 0.27 | 0.18 | 0.21 |
| Czech | 0.01 | 0.22 | 0.48 | 0.29 | 0.30 | 0.40 | 0.15 | 0.15 |
| Romanian | 0.08 | 0.24 | 0.43 | 0.25 | 0.48 | 0.21 | 0.17 | 0.14 |
| Persian | 0.07 | 0.20 | 0.45 | 0.28 | 0.29 | 0.14 | 0.20 | 0.36 |

**Fig 4. The prevalence of the four types of trajectories for bots and humans and for different language editions of Wikipedia.** The darker the shading of the cell, the higher the proportion for that type of trajectory for the language. Bot-bot interactions occur at a slower rate and are more balanced, in the sense that reverts go back and forth between the two bots. Further, bot-bot interactions are more balanced in smaller language editions of Wikipedia.

Looking at the prevalence of these four types of trajectories for bots and humans and across languages, we confirm the previous observations: bot-bot interactions occur at a slower rate and are more balanced, in the sense that reverts go back and forth between the two bots (Fig 4). Further, we find that bot-bot interactions are more balanced in smaller language editions of Wikipedia. This could be due to the fact that bots are more active in smaller editions and hence, interactions between them are more likely to occur. Less intuitively, however, this observation also suggests that conflict between bots is more likely to occur when there are fewer bots and when, common sense would suggest, coordination is easier.

# Discussion

Our results show that, although in quantitatively different ways, bots on Wikipedia behave and interact as unpredictably and as inefficiently as the humans. The disagreements likely arise from the bottom-up organization of the community, whereby human editors individually create and run bots, without a formal mechanism for coordination with other bot owners. Delving deeper into the data, we found that most of the disagreement occurs between bots that specialize in creating and modifying links between different language editions of the encyclopedia. The lack of coordination may be due to different language editions having slightly different naming rules and conventions.

In support of this argument, we also found that the same bots are responsible for the majority of reverts in all the language editions we study. For example, some of the bots that revert the most other bots include Xqbot, EmausBot, SieBot, and VolkovBot, all bots specializing in fixing inter-wiki links. Further, while there are few articles with many bot-bot reverts (S7 Fig), these articles tend to be the same across languages. For example, some of the articles most contested by bots are about Pervez Musharraf (former president of Pakistan), Uzbekistan, Estonia, Belarus, Arabic language, Niels Bohr, Arnold Schwarzenegger. This would suggest that a significant portion of bot-bot fighting occurs across languages rather than within. In contrast, the articles with most human-human reverts tend to concern local personalities and entities and tend to be unique for each language (26).

Our data cover a period of the evolution of Wikipedia when bot activity was growing. Evidence suggests that this period suddenly ended in 2013 (http://stats.wikimedia.org/EN/PlotsPngEditHistoryTop.htm). This decline occurred because at the beginning of 2013 many language editions of Wikipedia started to provide inter-language links via Wikidata, which is a collaboratively edited knowledge base intended to support Wikipedia. Since our results were largely dictated by inter-language bots, we believe that the conflict we observed on Wikipedia no longer occurs today. One interesting direction for future research is to investigate whether the conflict continues to persist among the inter-language bots that migrated to Wikidata.

Wikipedia is perhaps one of the best examples of a populous and complex bot ecosystem but this does not necessarily make it representative. As Table 1 demonstrates, we have investigated a very small region of the botosphere on the Internet. The Wikipedia bot ecosystem is gated and monitored and this is clearly not the case for systems of malevolent social bots, such as social bots on Twitter posing as humans to spread political propaganda or influence public discourse. Unlike the benevolent but conflicting bots of Wikipedia, many malevolent bots are collaborative, often coordinating their behavior as part of botnets (27). However, before being able to study the social interactions of these bots, we first need to learn to identify them (28).

Our analysis shows that a system of simple bots may produce complex dynamics and unintended consequences. In the case of Wikipedia, we see that benevolent bots that are designed to collaborate may end up in continuous disagreement. This is both inefficient as a waste of resources, and inefficacious, for it may lead to local impasse. Although such disagreements represent a small proportion of the bots' editorial activity, they nevertheless bring attention to the complexity of designing artificially intelligent agents. Part of the complexity stems from the common field of interaction – bots on the Internet, and in the world at large, do not act in isolation, and interaction is inevitable, whether designed for or not. Part of the complexity stems from the fact that there is a human designer behind every bot, as well as behind the environment in which bots operate, and that human artifacts embody human culture. As bots continue to proliferate and become more sophisticated, social scientist will need to devote more attention to understanding their culture and social life.

# Materials and Methods

## Data

Wikipedia is an ecosystem of bots. Some of the bots are "editing bots", that work on the articles. They undo vandalism, enforce bans, check spelling, create inter-language links, import content automatically, etc. Other bots are non-editing: these bots mine data, identify vandalism, or identify copyright violations.

In addition to bots, there are also certain automated services that editors use to streamline their work. For example, there are automated tools such Huggle and STiki, which produce a filtered set of edits to review in a live queue. Using these tools, editors can instantly revert the edit in question with a single click and advance to the next one. There are also user interface extensions and in-browser functions such as Twinkle, rollback, and undo, which also allow editors to revert with a single click. Another automated service that is relatively recent and much more sophisticated is the Objective Revision Evaluation Service (ORES). It uses machine-learning techniques to rank edits with the ultimate goal to identify vandals or low-quality contributions.

Our research focuses on editing bots. Our data contain who reverts whom, when, and in what article. To obtain this information, we analyzed the Wikipedia XML Dumps (https://dumps.wikimedia.org/mirrors.html) of 13 different language editions. To detect restored versions of an article, a hash was calculated for the complete article text following each revision and the hashes were compared between revisions (23). The data cover the period from the beginning of Wikipedia (January 15, 2001) until February 2, 2010 – October 31, 2011, the last date depending on when the data was collected for the particular language edition. This time period captures the "first generation" of Wikipedia bots, as in later years, Wikidata took over some of the tasks previously controlled by Wikipedia. The sample of languages covers a wide range of Wikipedia editions in terms of size; for example, it includes the four largest editions by number of edits and number of editors. In terms of cultural diversity, the sample covers a wide range of geographies.

Wikipedia requires that human editors create separate accounts for bots and that the bot account names clearly indicate the user is a bot, usually by including the word "bot" (https://en.wikipedia.org/wiki/Wikipedia:Bot_policy). Hence, to identify the bots, we selected all account names that contain different spelling variations of the word "bot." We supplemented this set with all accounts that have currently active bot status in the Wikipedia database but that may not fit the above criterion (using https://en.wikipedia.org/wiki/Wikipedia:Bots/Status as of August 6, 2015). We thus obtained a list of 6,627 suspected bots.

We then used the Wikipedia API to check the "User" page for each suspected bot account. If the page contained a link to another account, we confirmed that the current account was a bot and linked it to its owner. For pages that contained zero or more than one links to other accounts, we manually checked the "User" and "User_talk" pages for the suspected bot account to see if it is indeed a bot and to identify its owner. The majority of manually checked accounts were vandals or humans, so we ended up with 1,549 bots, each linked to its human owner.

We additionally labeled human editors as vandals if they had all their edits reverted by others. This rule meant that we labeled as vandals also newcomers who became discouraged and left Wikipedia after all their initial contributions were reverted. Since we are interested in social interactions emerging from repeated activity, we do not believe that this decision affects our results.

Using the revert data, we created a directed two-layer multi-edge network, where ownership couples the layer of human editors and the layer of bots (29). To build the network, we assumed that a link goes from the editor who restored an earlier version of the article (the "reverter") to the editor who made the revision immediately after that version (the "reverted"). All links were time-stamped. We collapsed multiple bots to a single node if they were owned by the same human editor; these bots were usually accounts for different generations of the same bot with the same function. In the network, reverts can be both intra- and inter-layer: they occur within the human layer, within the bot layer, and in either direction between the human and bot layers. The multi-layer network was pruned by removing self-reverts, as well as reverts between a bot and its owner.

## Interaction trajectories

We model the interaction trajectories in two-dimensional space, where the x-axis measures time and the y-axis measures the difference between the number of times $i$ has reverted $j$ and the number of times $j$ has reverted $i$. To construct the trajectories, starting from $y_0 = 0$, $y_t = y_{t-1} + 1$ if $i$ reverts $j$ at time $t$ and $y_t = y_{t-1} - 1$ if $j$ reverts $i$ at time $t$; the labels $i$ and $j$ are assigned so that $y >= 0$ for the majority of the $ij$ interaction time. We analyze three properties of the trajectories:

- Latency. We define latency as the mean log time in seconds between successive reverts: $\mu(\log_{10} \Delta t)$.

- Imbalance. We define imbalance as the final proportion of reverts between $i$ and $j$ that were not reciprocated: $|r_i - r_j| / (r_i + r_j)$, where $r_i$ and $r_j$ are the number of times $i$ reverted $j$ and $j$ reverted $i$, respectively.

- Reciprocity. We define reciprocity as the proportion of observed turning points out of all possible: (# turning points) / $(r_i + r_j - 1)$ , where $r_i$ and $r_j$ are the number of times $i$ reverted $j$ and $j$ reverted $i$, respectively. A turning point occurs when the user who reverts at time $t$ is different from the user who reverts at time $t$+1.

## K-means clustering

To identify the number of clusters $k$ that best represents the data, we apply the elbow and silhouette methods on trajectories of different minimum length. The rationale behind restricting the data to long trajectories only is that short trajectories tend to have extreme values on the three features, thus possibly skewing the results. According to the elbow method, we would like the smallest $k$ that most significantly reduces the sum of squared errors for the clustering. According to the silhouette method, we would like the $k$ that maximizes the separation distance between clusters and thus gives us the largest silhouette score.

Although the elbow method suggests that four clusters provide the best clustering, the silhouette method indicates that the data cannot be clustered well (S8 Fig). We do not necessarily expect that trajectories cluster naturally; rather, we employ clustering in order to quantify the differences between the interactions of bots versus humans across languages. We hence analyze the clustering with $k = 4$. This clustering also has the advantage of yielding four types of trajectories that intuitively make sense.

# Acknowledgements

The authors thank Wikimedia Deutchland e.V. and Wikimedia Foundation for the live access to the Wikipedia data via Toolserver. The data reported in the paper are available at http://wwm.phy.bme.hu.

# Supporting Information

**S1 Table. Descriptive statistics for the bot-bot layer and the human-human layer in the multi-layer networks of reverts.**

| | Number of nodes | Avg. number reverts per node | Reverts / edits | Prop. dyads with at least one revert reciprocated | Assortativity by number of edits | Avg. clustering | Avg. clustering / avg. clustering in random network |
|---|---|---|---|---|---|---|---|
| **Bot-bot** | | | | | | | |
| English | 319 | 104.6 | 0.002 | 0.46 | -0.02 | 0.43 | 19 |
| Japanese | 182 | 100.4 | 0.006 | 0.57 | -0.13 | 0.58 | 9 |
| Spanish | 204 | 71.7 | 0.003 | 0.53 | -0.06 | 0.57 | 12 |
| French | 225 | 59.3 | 0.004 | 0.47 | -0.1 | 0.5 | 12 |
| Portuguese | 164 | 185 | 0.006 | 0.57 | -0.12 | 0.64 | 8 |
| German | 178 | 24.1 | 0.004 | 0.43 | -0.1 | 0.4 | 10 |
| Chinese | 151 | 103 | 0.006 | 0.59 | -0.16 | 0.62 | 9 |
| Hebrew | 124 | 83.9 | 0.006 | 0.59 | -0.11 | 0.59 | 7 |
| Hungarian | 116 | 66.8 | 0.004 | 0.54 | -0.13 | 0.6 | 8 |
| Czech | 122 | 59 | 0.005 | 0.57 | -0.18 | 0.56 | 8 |
| Arabic | 132 | 161.7 | 0.011 | 0.6 | -0.05 | 0.6 | 7 |
| Romanian | 104 | 70.8 | 0.005 | 0.55 | -0.11 | 0.6 | 7 |
| Persian | 106 | 63.8 | 0.005 | 0.5 | -0.05 | 0.53 | 8 |
| **Human-human** | | | | | | | |
| English | 4127880 | 3.1 | 0.079 | 0.09 | -0.05 | 0.04 | 72370 |
| Japanese | 193203 | 2.6 | 0.026 | 0.16 | -0.05 | 0.02 | 2971 |
| Spanish | 508815 | 2.3 | 0.070 | 0.09 | -0.11 | 0.1 | 33433 |
| French | 181395 | 2.6 | 0.045 | 0.09 | -0.02 | 0.04 | 4011 |
| Portuguese | 262293 | 2.3 | 0.066 | 0.09 | -0.14 | 0.12 | 19762 |
| German | 206734 | 2.2 | 0.069 | 0.11 | -0.1 | 0.03 | 3703 |
| Chinese | 66470 | 3.2 | 0.028 | 0.18 | -0.14 | 0.08 | 3377 |
| Hebrew | 70816 | 2.9 | 0.047 | 0.13 | -0.13 | 0.2 | 7458 |
| Hungarian | 21036 | 2.4 | 0.016 | 0.11 | -0.13 | 0.1 | 1265 |
| Czech | 23792 | 3.1 | 0.035 | 0.1 | -0.18 | 0.19 | 2262 |
| Arabic | 39083 | 2.2 | 0.044 | 0.08 | -0.17 | 0.11 | 2947 |
| Romanian | 16625 | 2.1 | 0.027 | 0.1 | -0.2 | 0.11 | 1371 |
| Persian | 18657 | 3.6 | 0.056 | 0.16 | -0.13 | 0.21 | 1972 |

Bots revert each other to a great extent. They also reciprocate each other's reverts to a considerable extent. Their interactions are not as clustered as for human editors. Still, both for bots and humans, more senior editors tend to revert less senior editors, as measured by node assortativity by number of edits completed.

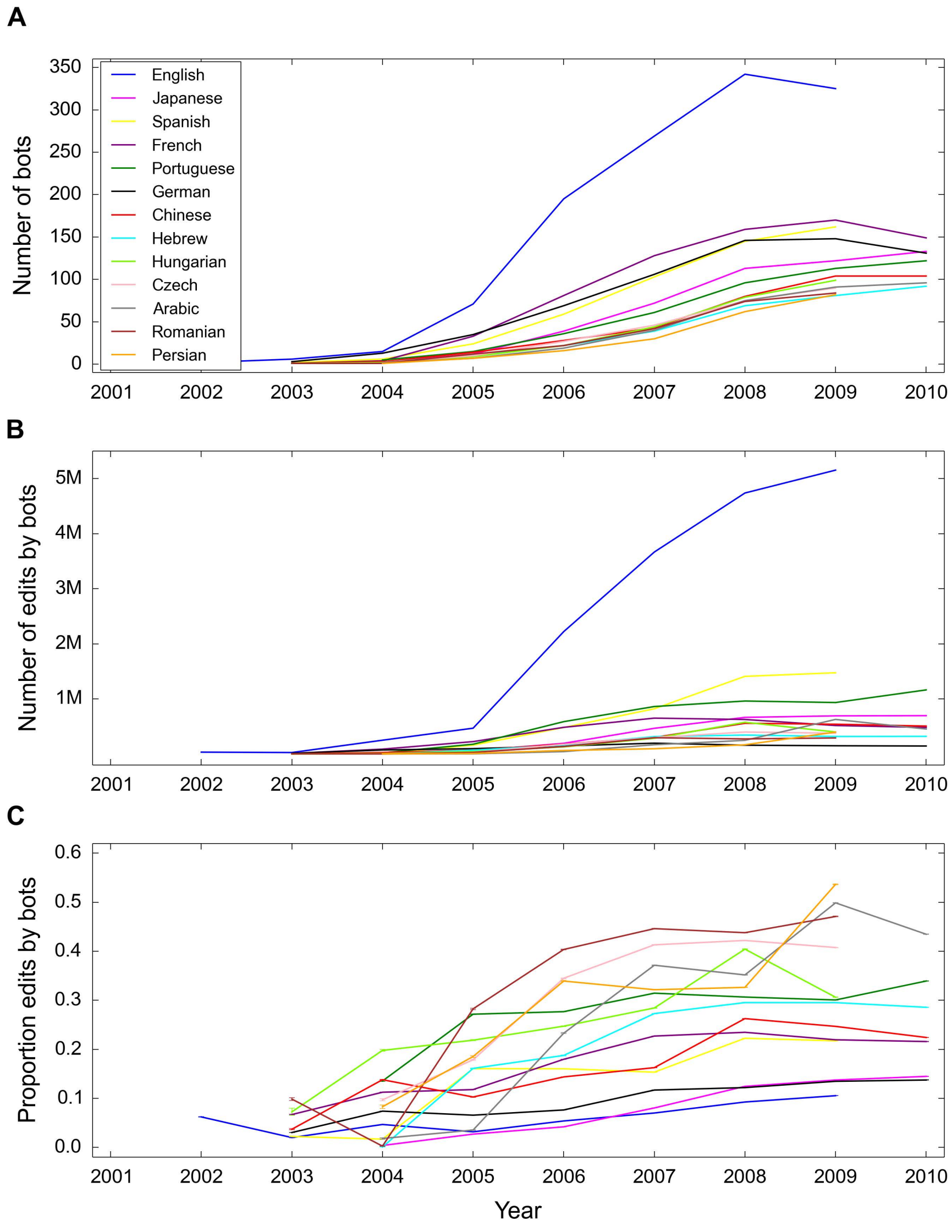

**S1 Fig. The number of bots, the number of edits by bots, and the proportion of edits done by bots between 2001 and 2010.** Between 2003 and 2008 the number of bots and their activity have been increasing. This trend, however, appears to have subsided after 2008, suggesting that the system may have stabilized.

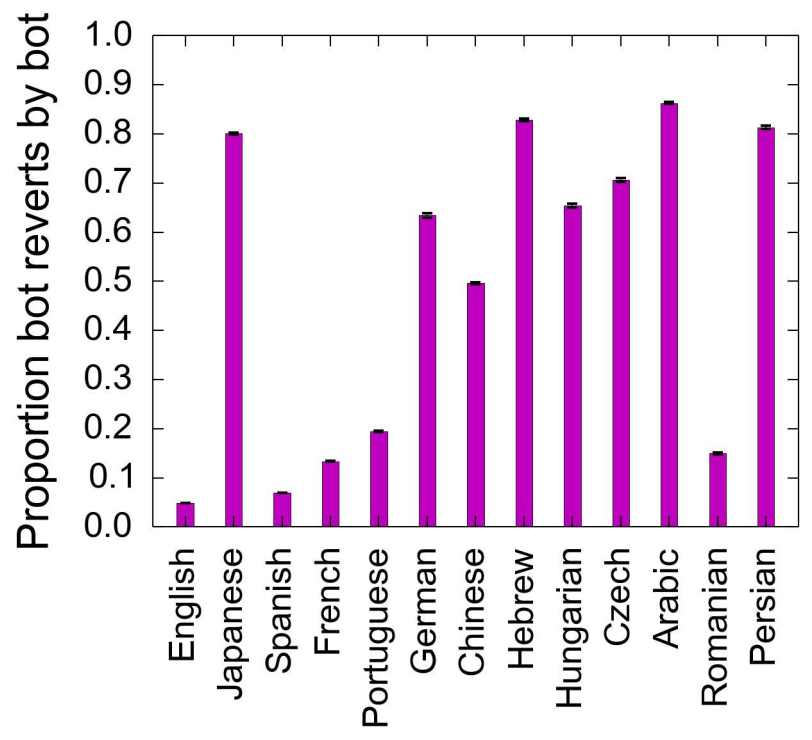

**S2 Fig. For the majority of languages, bots are mainly reverted by other bots, as opposed to human editors or vandals.** English and the Romance languages in our data present exceptions, with less than 20% of bot reverts are done by other bots.

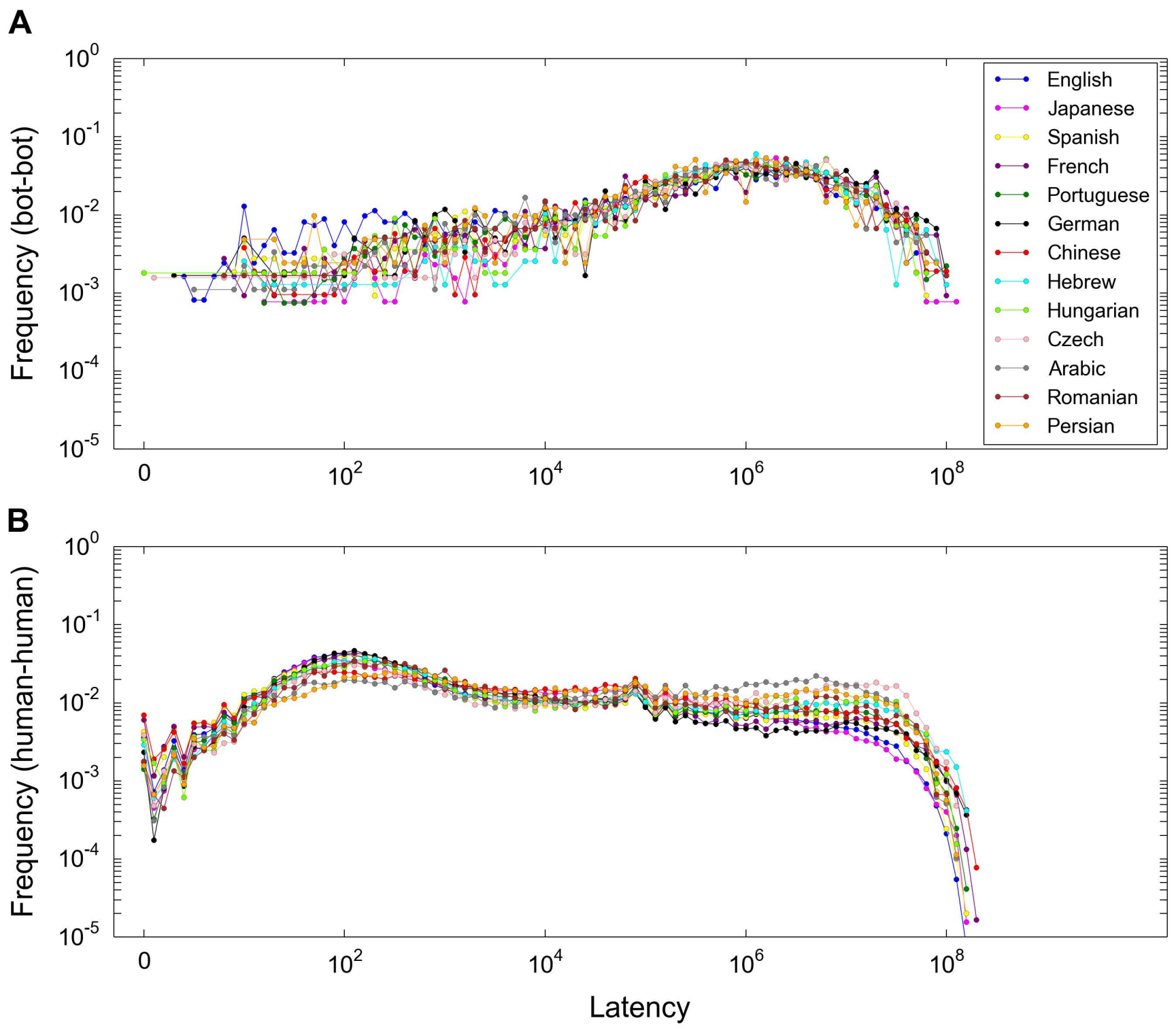

**S3 Fig. Bot-bot interactions have different characteristic time scale than human-human interactions.** The figures show the distribution of interactions for a particular latency, where we define latency as the mean log time in seconds between successive reverts. (A) Bot-bot interactions have a characteristic latency of 1 month, as indicated by the peak in the figure. (B) Human-human interactions occur with a latency of 2 minutes, 24 hours, or 1 year.

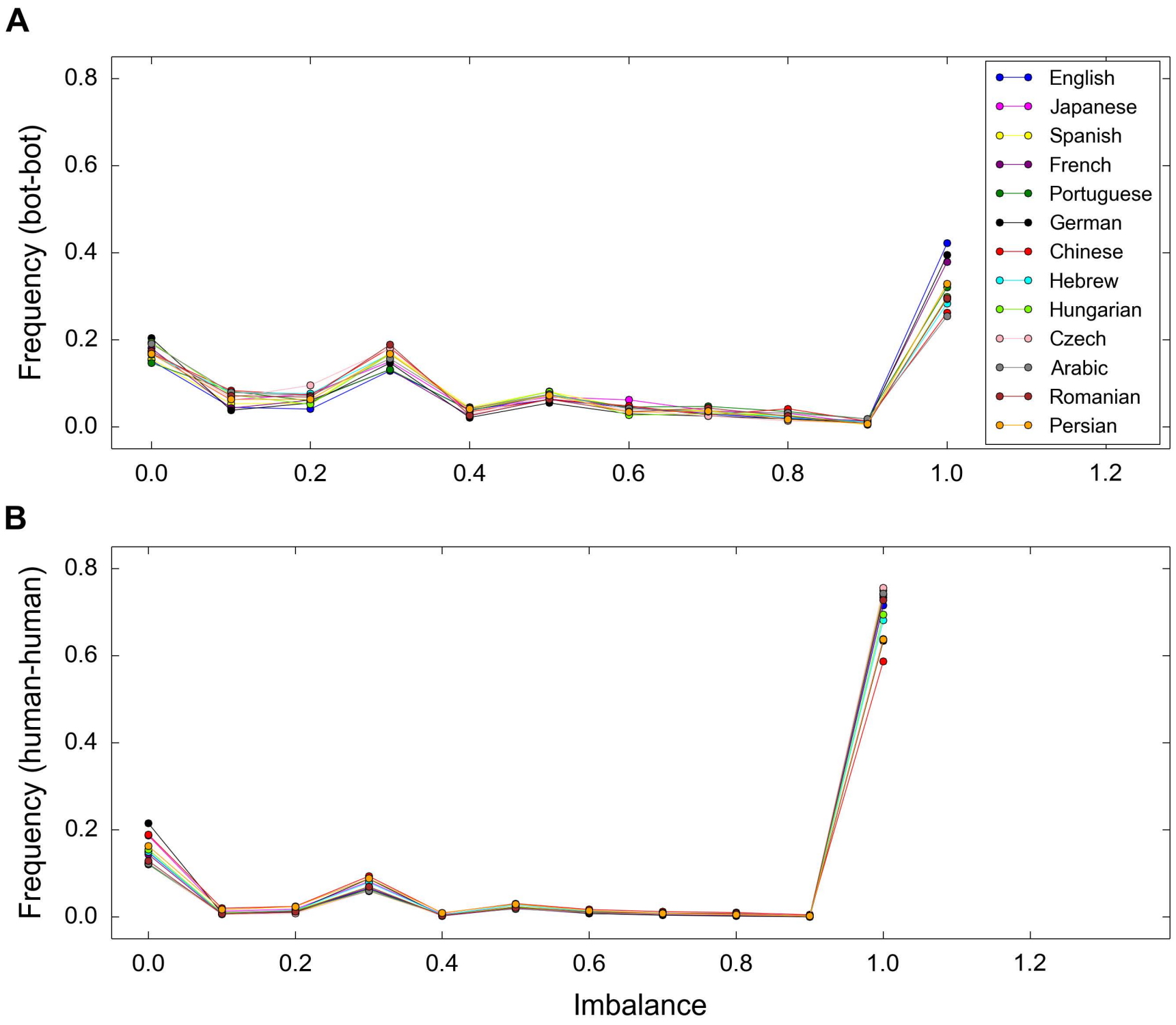

**S4 Fig. Bot-bot interactions are on average more balanced than human-human interactions.** We define imbalance as the final proportion of reverts between *i* and *j* that were not reciprocated. (A) A significant proportion of bot-bot interactions have low imbalance. (B) The majority of human-human interactions are perfectly unbalanced.

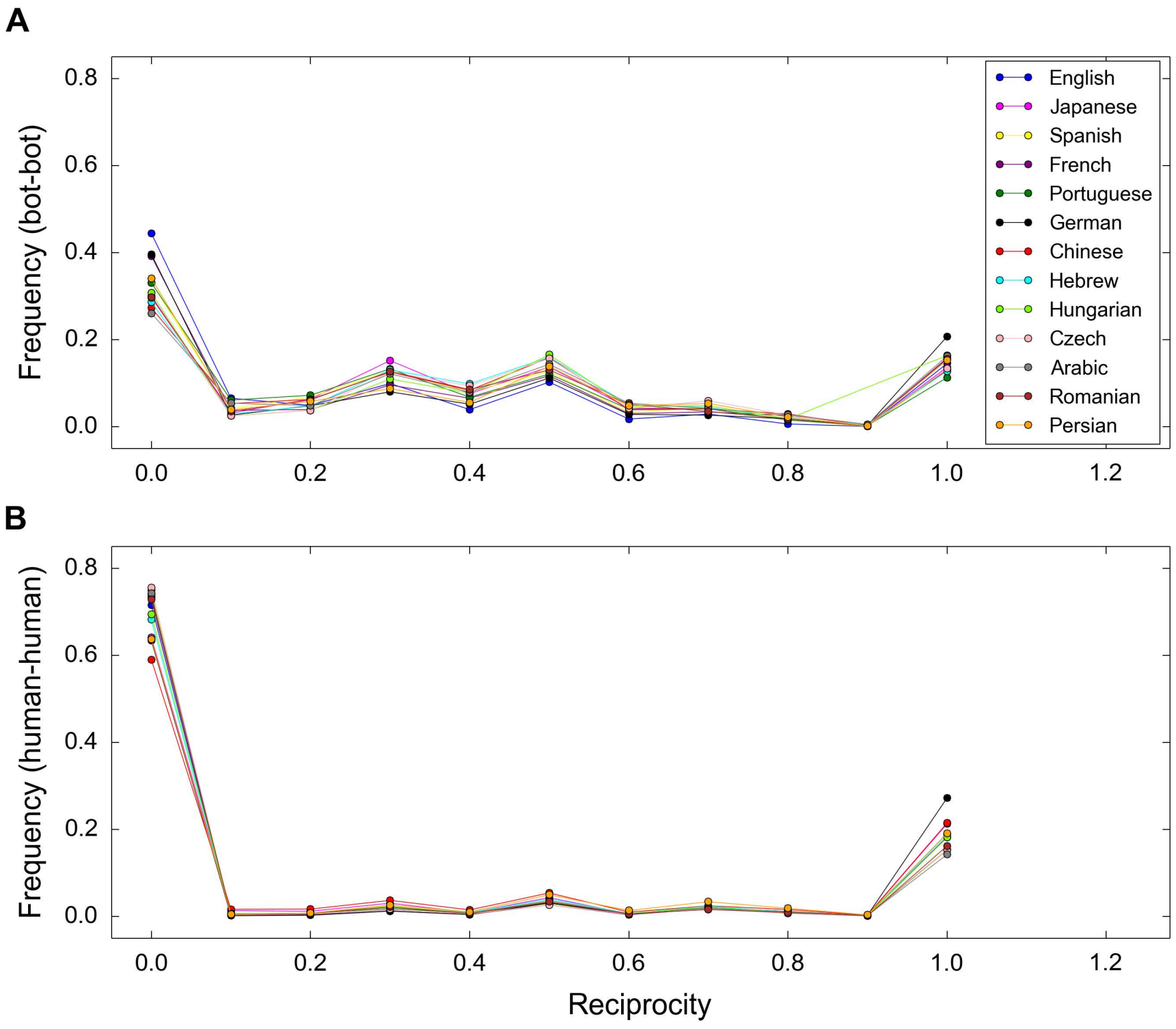

**S5 Fig. Bots reciprocate much more than humans do also at a smaller timescale.** We measure reciprocity as the proportion of observed turning points out of all possible. (A) A significant proportion of bot-bot interactions have intermediate or high values of reciprocity. (B) The majority of human-human interactions are not reciprocated.

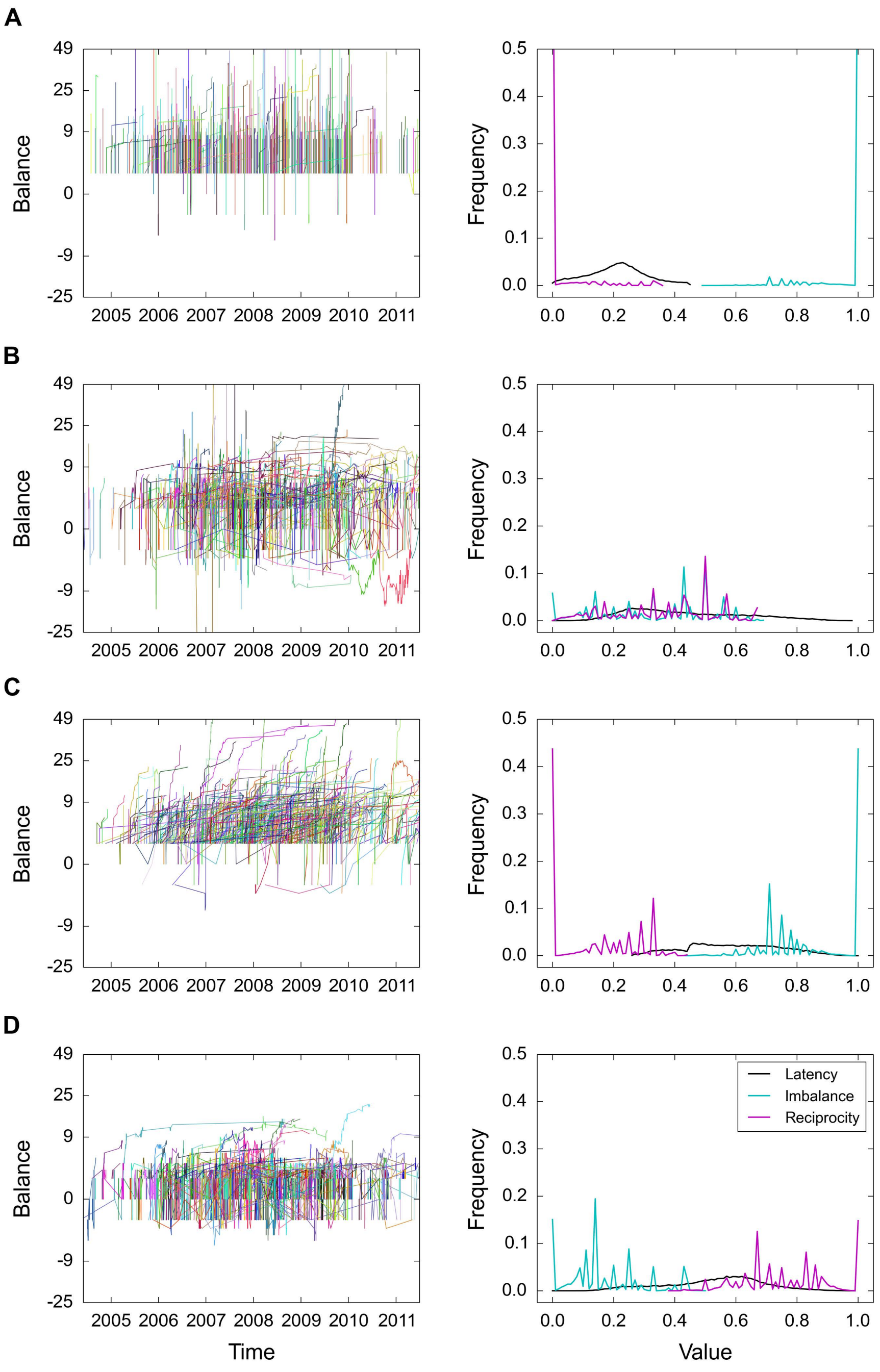

**S6 Fig. Four types of interaction trajectories suggested by the k-means analysis.** The left panels show a sample of the trajectories, including bot-bot and human-human interactions and trajectories from all languages. The right panels show the distribution of latency, imbalance, and reciprocity for each type of trajectory. The three properties measure the average steepness, the *y*-value of the last point, and the jaggedness of the trajectory, respectively. (A) Fast unbalanced trajectories have low reciprocity and latency and high imbalance. (B) Somewhat balanced trajectories have intermediate imbalance and reciprocity. (C) Slow unbalanced trajectories have low reciprocity and high latency and imbalance. (D)

Well balanced trajectories have low imbalance and high reciprocity.

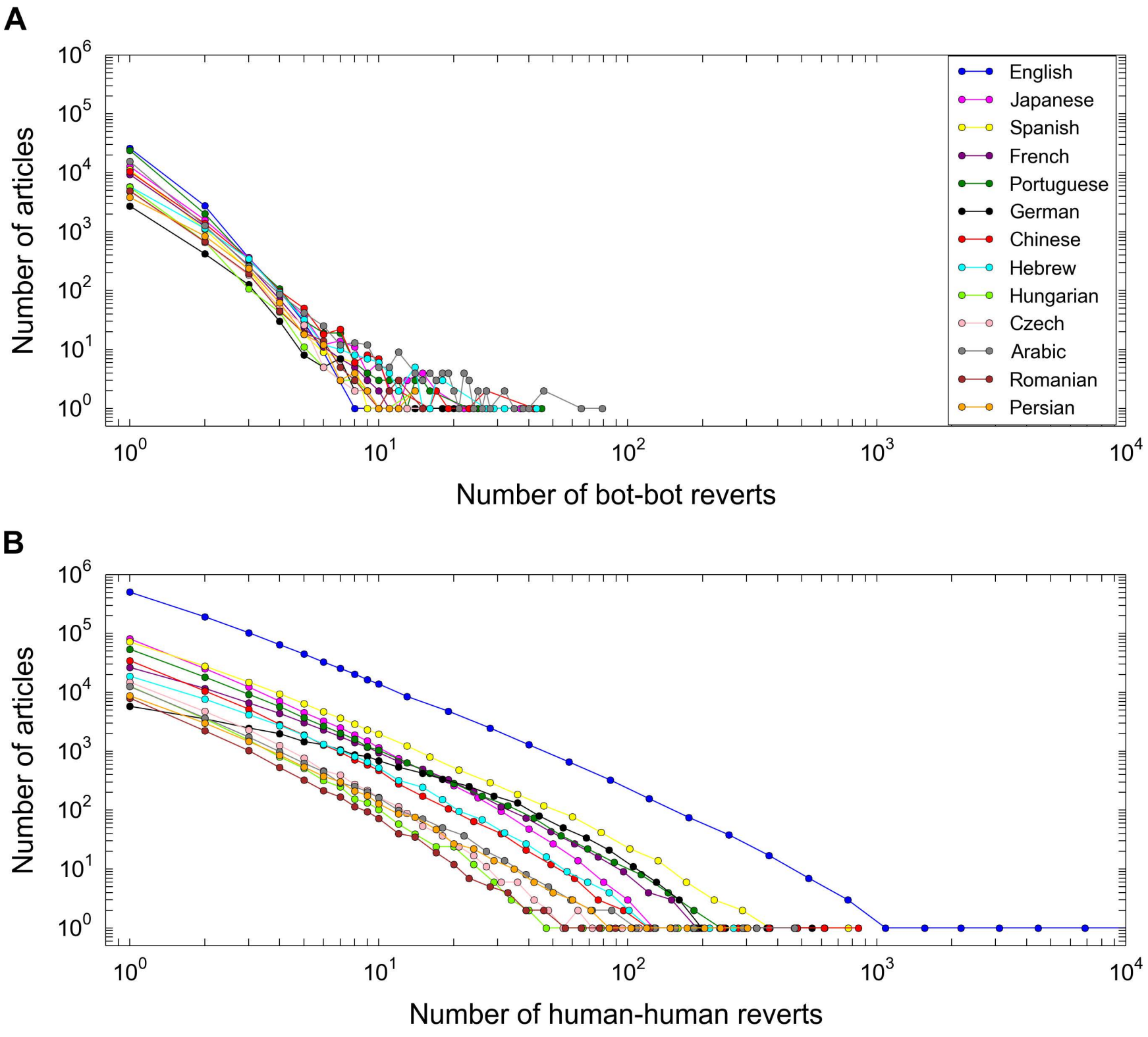

**S7 Fig. The number of articles with a certain number of bot-bot and human-human reverts.** (A) Few articles include more than 10 bot-bot reverts. The most contested articles tend to be about foreign countries and personalities. Further, the same articles also re-appear in different languages. (B) There are many articles that are highly contested by humans. The most contested articles tend to concern local personalities and entities. It is rare that a highly contested article in one language will be also highly contested in another language.

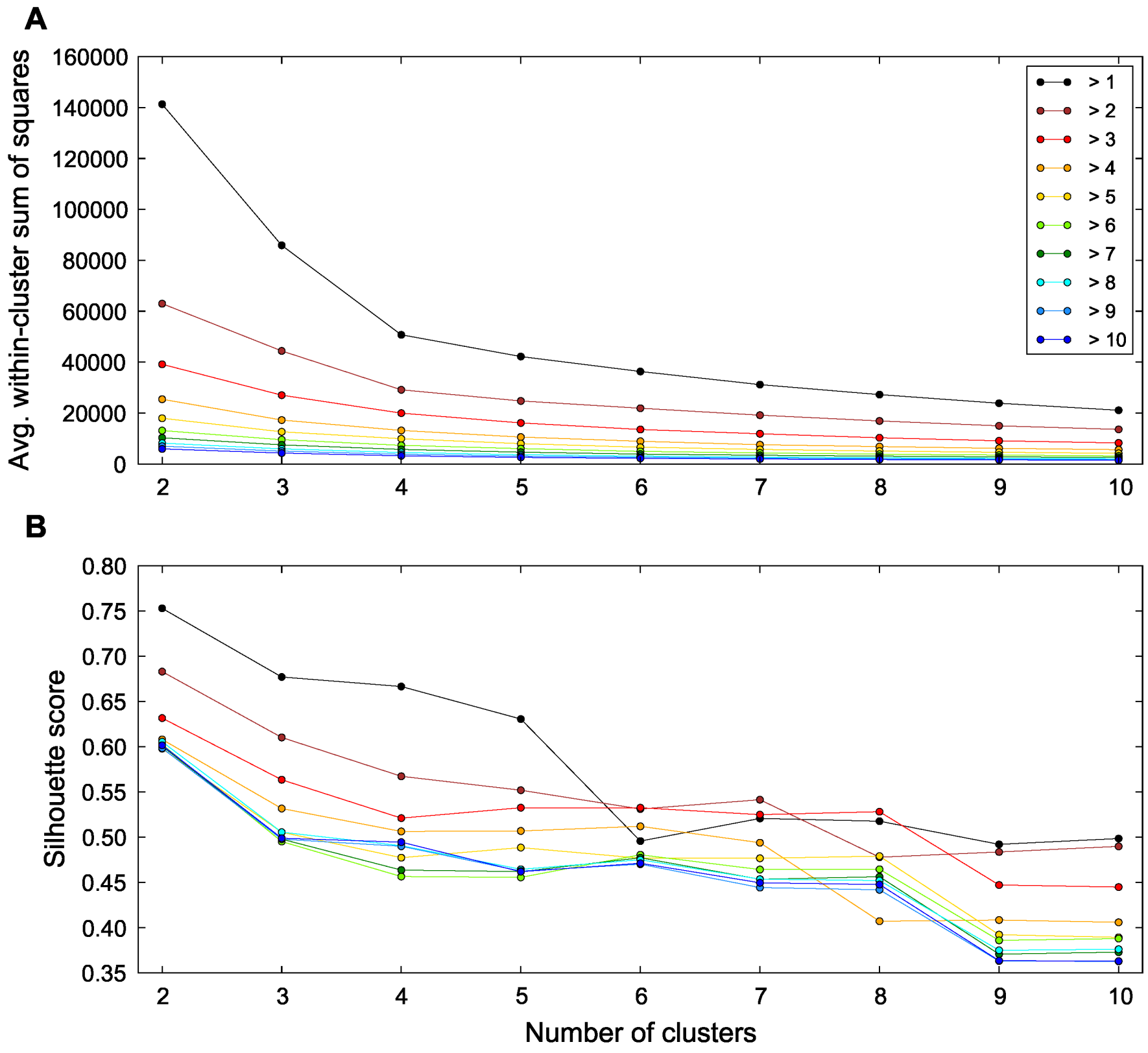

**S8 Fig. Performance of the k-means clustering algorithm for different number of clusters and for sub-samples with different minimum length of trajectories.** (A) The elbow method requires the smallest $k$ that most significantly reduces the sum of squared errors for the clustering. Here, the method suggests that four clusters give the best clustering of the data. (B) The silhouette method requires the $k$ that maximizes the separation distance between clusters, i.e. the largest silhouette score. Here, the method suggests that the clustering performs worse as the number of clusters increases.